\documentclass{ws-procs9x6}

\begin{document}

\title{Search for new physics in $B_s$-mixing}

\author{Alexander J. Lenz$^*$}

\address{Faculty of Physics, University of Regensburg,\\
D-93040 Regensburg, Germany\\
$^*$E-mail: alexander.lenz@physik.uni-regensburg.de
}

\begin{abstract}
We present the current status of the search for new physics
effects in the mixing quantities $\Delta M_s$, $\Delta \Gamma_s$ and 
$\phi_s$ of the neutral $B_s$-system.
\end{abstract}


\bodymatter
\section{Introduction}\label{aba:sec1}
Despite the enormous success of the standard model there
is still room for new physics to be detected at currently running 
experiments. Huge efforts have been made in recent years in the precision measurement
and precision calculation of flavor physics observables at the B-factories and at TeVatron, 
see e.g. Ref.~[\refcite{Battaglia:2003in}] for a review and references therein.
The system of the neutral $B_s$ mesons seems to be particular promising to find hints for 
new physics
(for a recent review of B-mixing see Ref.~[\refcite{Schneider:2008dy}]): the standard model 
contribution is suppressed strongly, so even small new physics contributions might be of 
comparable size
and the hadronic uncertainties are under good control.
\\
In the standard model the mixing of neutral B-meson ist described, by the
box diagrams, see e.g. [\refcite{TeVatron,Bigi:2000yz,Lenz:2008xt,Lenz:2007eu, Lenz:2007nk,Lenz:2006ni}] 
for more details. 
The absorptive part $\Gamma_{12}$ of the box diagrams is sensitive light internal particles
and the dispersive part $M_{12}$ is sensitive to heavy internal particles.
The two complex quantities $M_{12}$ and $\Gamma_{12}$ can be related to the following
physical quantities:
\begin{itemize}
\item The mass difference $\Delta M_s$ between the heavy and the light mass eigenstates 
      of the neutral $B$ mesons:
      \begin{equation}
      \Delta M_s = 2 | M_{12}|\, .
      \end{equation}
\item The decay rate difference $\Delta \Gamma_s$ between the heavy and the light mass eigenstates 
      of the neutral $B$ mesons:
      \begin{equation}
      \Delta \Gamma_s = 2 | \Gamma_{12}| \cos \left( \phi_s \right)
      \end{equation}
      with the weak phase $\phi_s := $Arg$ \left( -M_{12}/\Gamma_{12} \right)$.
\item The tiny CP asymmetries in semileptonic $B$-decays $a_{sl}^s$
      \begin{equation}
      a_{sl}^s = \frac{| \Gamma_{12}|}{|M_{12}|} \sin \left( \phi_s \right) \, .
      \end{equation}
\end{itemize}
For the weak phase $\phi_s$ different notations are used in the literature, which
led already to some confusion. For more details on the definitions see the
{\it Note added} in [\refcite{Lenz:2007nk}].
\\
Recently there were several claims of possible new physics effects in the $B_s$-mixing system
in the literature:
\begin{enumerate}
\item End of 2006 a 2 $\sigma$-deviation was found \cite{Lenz:2006hd}, if all mixing
      quantities in the $B_s$-system were combined.
\item This was more or less confirmed in july 2007 by UT-Fit \cite{Bona:2007vi}.
\item With new data available UT-fit \cite{Bona:2008jn} claimed in march 2008 
      a 3.7 $\sigma$-deviation from the standard model. Since from the experiments (D0 and CDF)
      the full information about the likelihoods was not available at that time, the combination of
      the data in Ref.~[\refcite{Bona:2008jn}] had to rely on some assumptions.
\item This analysis is currently redone - with the missing experimental information - 
      by CKM Fitter in collaboration with the authors of 
      Ref.~[\refcite{Lenz:2006hd}]
      \cite{CKMFITTER}, preliminary results \cite{CKMFITTERprelim} show a deviation of less 
      than 3 $\sigma$.
\end{enumerate}
The above claims are based on the following experimental data for the $B_s$ mixing system, mostly  
from D0 and CDF:
\begin{itemize}
\item The mass difference $\Delta M_s$ was measuered at CDF \cite{DeltaMCDF} and at D0 
      \cite{DeltaMD0} and the numbers were combined from HFAG \cite{HFAG} to
      \begin{eqnarray}
      \Delta M_s & = & 17.78 \pm 0.12 \, \mbox{ps}^{-1} \, . 
      \end{eqnarray}
\item D0 \cite{taggedD0} and CDF \cite{taggedCDF} performed a tagged analysis of 
      the decay $B_s \to J/\Psi \phi$ to determine the decay rate difference $\Delta \Gamma_s$ and the
      weak mixing angle $\phi_s$. HFAG \cite{HFAG} combines the values to, see Fig. (\ref{HFAG1})
      \begin{eqnarray}
      \Delta \Gamma_s & = & 0.154^{+0.054}_{-0.070}\, \mbox{ps}^{-1} \, ,
      \\
      \phi_s          & = & -0.77^{+0.29}_{0.37} \, .
      \end{eqnarray}

\begin{figure}
\begin{center}
\psfig{file=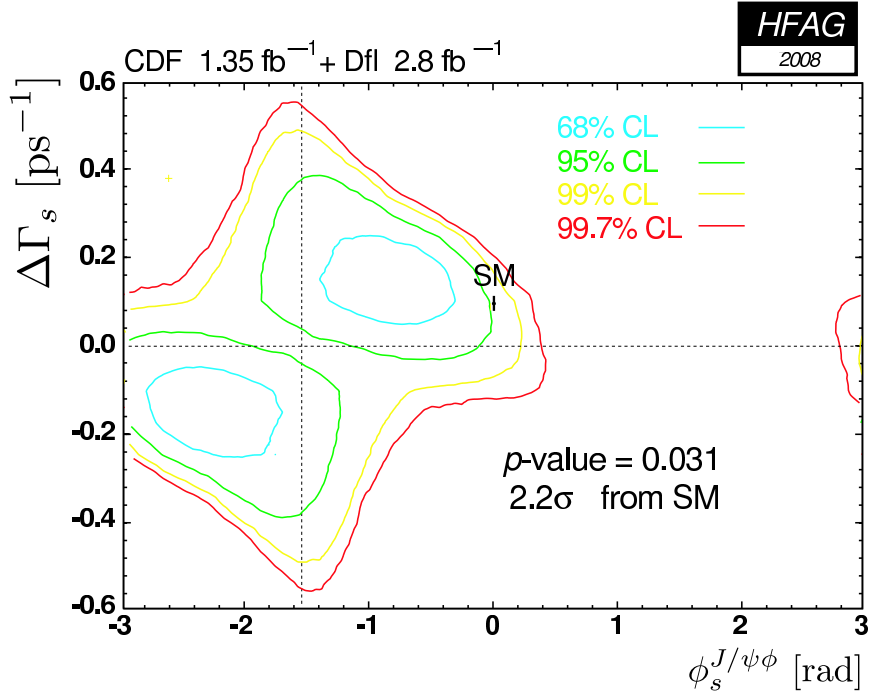,width=4.5in}
\end{center}
\caption{The combined experimental values for $\Delta \Gamma_s$ and $\phi_s$ 
from the tagged analysis of the decay $B_s \to J/\Psi \phi$ from D0 and CDF.
}
\label{HFAG1}
\end{figure}
      The result from CDF \cite{taggedCDF} is now superseeded by Ref.~[\refcite{taggedCDFnew}].
\item The semileptonic CP asymmetry can be obtained from the dimuon asymmetry
      (CDF \cite{dimuonCDF}, D0 \cite{dimuonD0}) or it can be measure directly
      (D0 \cite{aslD0un}).
      These numbers were combined from HFAG \cite{HFAG} to
      \begin{eqnarray}
      a_{sl}^s & = & + 0.0016 \pm 0.0085 \, .
      \end{eqnarray}
      The untagged result from Ref.~[\cite{aslD0un}] is now superseeded by the new tagged result 
      \cite{aslD0} 
      \begin{eqnarray}
      a_{sl}^s & = & - 0.0024 \pm 0.0117 ^{+0.0015}_{-0.0024} \, .
      \end{eqnarray}
      \end{itemize}
There are numerous applications of new physics models to the $B_s$ mixing sector, 
for some recent examples, see e.g. Refs. [
\refcite{Altmannshofer:2008hc,Chen:2008ug,Soni:2008bc,Lee:2008xr,Botella:2008qm,Buras:2008nn,
      Kifune:2007fj,Cabarcas:2007my,Chen:2007cz,
      Joshipura:2007cs,Lenz:2007nj,Badin:2007bv,Lunghi:2007ak,Chen:2007dg,Joshipura:2007sf,
      Dighe:2007gt,Chen:2007nx,Parry:2007fe,Parry:2008sr}].
TeVatron is continuing to take data and we will get more precise data from the upcomming
experiments at LHC \cite{Buchalla:2008jp} or possibly at a SuperB-factory \cite{Bona:2007qt}
running also at the $\Upsilon (5s)$-resonance.
\section{Strategy to search for new physics}
\begin{figure}
\begin{center}
\psfig{file=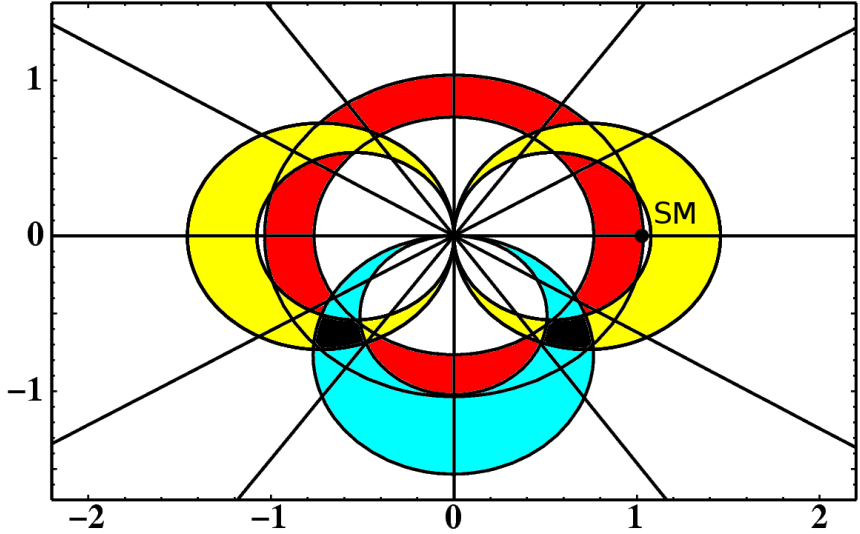,width=4.5in}
\end{center}
\caption{The bounds in the complex $\Delta$-plane, obtained by comparing experiment and theory
for the mixing quantities. The red circle comes from $\Delta M_s$, the yellow band from
$\Delta \Gamma_s / \Delta M_s$, the light blue range from the semileptonic CP asymmetries
and the rays through the origin from a direct determination of $\phi_s$.
}
\label{Bound1}
\end{figure}

In [\refcite{Lenz:2006hd}] we worked out a model independent analysis of new physics effects in 
$B$-mixing.
$\Gamma_{12}$ is due to real intermediate states, i.e. particles which are lighter than $m_B$. 
Any new physics contributions to $\Gamma_{12}$ affects also tree-level $B$-decays. Since no evidence
for sizeable new physics effects in tree-level $B$-decays has been found so far, it reasonable to
assume that $\Gamma_{12}$ is described by the standard model contributions alone. Deviations from 
that assumption are expected to be smaller than the hadronic unvertainties in the standard model
prediction for $\Gamma_{12}$. $M_{12}$, however, might be affected by large new physics effects. 
We write therefore
\begin{eqnarray}
M_{12}^s & = & M_{12}^{\it SM,s} \cdot \Delta =  M_{12}^{\it SM,s} \cdot |\Delta| \cdot e^{i \phi_s^\Delta} 
\, ,
\\
\Gamma_{12}^s & = & \Gamma_{12}^{\it SM,s} \, ,
\end{eqnarray}
where all new physics effects are parameterized by the complex number $\Delta$.
Now we can relate the experimental observables in the mixing system with
the standard model predictions \cite{Lenz:2006hd} and with  $\Delta$.
\begin{eqnarray}
\Delta M_s  & = & \Delta M_s^{\rm SM} \,  |\Delta_s|
\nonumber
\\
& = &
(19.30 \pm 6.74 ) \, \mbox{ps}^{-1} \cdot | \Delta_s| \, ,
\label{bounddm}
\\
\Delta \Gamma_s  & = & 2 |\Gamma_{12}^s|
     \, \cos \left( \phi_s^{\rm SM} + \phi^\Delta_s \right)
\nonumber
\\
& = &(0.096 \pm 0.039) \, \mbox{ps}^{-1}
\cdot \cos \left( \phi_s^{\rm SM} + \phi^\Delta_s \right) \, ,
\label{bounddg}
\\
\frac{\Delta \Gamma_s}{\Delta M_s}
&= &
 \frac{|\Gamma_{12}^s|}{|M_{12}^{\rm SM,s}|}
\cdot \frac{\cos \left( \phi_s^{\rm SM} + \phi^\Delta_s \right)}{|\Delta_s|}
\nonumber
\\
& =&
\left( 4.97 \pm 0.94 \right) \cdot 10^{-3}
\cdot \frac{\cos \left( \phi_s^{\rm SM} + \phi^\Delta_s \right)}{|\Delta_s|} \, ,
\label{bounddgdm}
\\
a_{\rm fs}^s
&= &
 \frac{|\Gamma_{12}^s|}{|M_{12}^{\rm SM,s}|}
\cdot \frac{\sin \left( \phi_s^{\rm SM} + \phi^\Delta_s \right)}{|\Delta_s|}
\nonumber
\\
& = &\left( 4.97 \pm 0.94 \right) \cdot 10^{-3}
\cdot \frac{\sin \left( \phi_s^{\rm SM} + \phi^\Delta_s
  \right)}{|\Delta_s|} \, ,
\label{boundafs} \\
\lefteqn{\mbox{with}
 \qquad \phi_s^{\rm SM} = (4.2\pm 1.4) \cdot 10^{-3} \, .} 
\end{eqnarray}
By comparing experiment and theory, we can give bounds in the complex $\Delta$-plane 
\footnote{The bounds in the complex $\Delta$-plane are much more descriptive than 
in the $|\Delta|$-$\phi_s^\Delta$-plane, which is used also in the literature.
}. If nature would be such, that $\Delta$ has the values:
\begin{equation}
|\Delta| = 0.9 \, , \hspace{1cm} \phi_s^\Delta = \frac{\pi}{4} \, ,
\end{equation}
one would get the bounds shown in Fig. (\ref{Bound1}).
\begin{itemize}
\item $\Delta M_s                $ gives a bound on the absolute value of $\Delta$ 
                                   (c.f. Eq. (\ref{bounddm})), 
                                 which is represented by the red band in Fig. (\ref{Bound1}).
\item $\phi_s^\Delta$ can be obtained directly from the angular analysis of the decay 
                                 $B_s \to J/ \Psi \phi$. With a considerably worse accuracy this
                                 phase can also be obtained from $\Delta \Gamma_s$ (c.f. Eq. (\ref{bounddg})).
\item $\Delta \Gamma_s / \Delta M_s$ gives a bound on $
                                   \cos \left( \phi_s^{\rm SM} + \phi^\Delta_s \right) / |\Delta|$ 
                                   (c.f. Eq. (\ref{bounddgdm})), which is represented by the
                                   yellow band in Fig. (\ref{Bound1}).
\item $a_{\rm fs}^s              $ gives a bound on $
                                   \sin \left( \phi_s^{\rm SM} + \phi^\Delta_s \right) / |\Delta|$ 
                                 (c.f. Eq. (\ref{boundafs})), which is represented by the
                                   light blue band in Fig. (\ref{Bound1}).
\end{itemize}
The overlap of all these bounds gives the values for Re($\Delta$) and Im($\Delta$).
Within the standard model one has Re($\Delta$)=1 and Im($\Delta$)=0.
\section{Theoretical framework and uncertainties}
In order to fulfill the above described program it is mandatory to have sufficient
control over the theoretical uncertainities in the standard model predictions.
\\
Inclusive decays can be described by the Heavy Quark Expansion (HQE) 
\cite{HQE1,HQE2,HQE3,HQE4,HQE5,HQE6,HQE7,HQE8},
for some recent examples see [\refcite{Lenz:1997aa,Lenz:1998qp,Lenz:2000kv}].
According to the HQE an inclusive  decay rate can be expanded in inverse powers of the heavy $b$-quark 
mass
\begin{equation}
\Gamma = \Gamma_0 + \left( \frac{\Lambda_{QCD}}{m_b} \right)^2 \Gamma_2
                  + \left( \frac{\Lambda_{QCD}}{m_b} \right)^3 \Gamma_3
                  + \left( \frac{\Lambda_{QCD}}{m_b} \right)^4 \Gamma_4
                  + ... \, .
\label{HQEformula}
\end{equation}
In order to estimate the theoretical accuracy for the 
mixing quantities $\Gamma_{12}$ and $M_{12}$, one first has to investigate
the general validity of the expansion in Eq.(\ref{HQEformula}).
This was done many times in the literature under the name of 
{\it violations of quark-hadron duality}, see e.g. [\refcite{Bigi:1998kc}] 
and references therein.
We follow a pragmatic strategy, as described in more detail in [\refcite{Lenz:2008xt}]:
the calculation of the mixing quantities $\Gamma_{12}$ is identical to the ones
of the  lifetimes, which are also known to NLO-QCD \cite{Beneke:2002rj,Franco:2002fc}.
Since experiment and the HQE predicition agree very well \cite{Lenz:2008xt}, we see
no room for sizeable violations of quark-hadron duality.
\\
All $\Gamma_i$s in Eq.(\ref{HQEformula}) are products of perturbatively calculable
Wilson coefficients and of non-perturbative matrix elements.
To be sure to achieve a reasonable theoretical accuracy we have to calculate up to a
sufficient order in the HQE and in QCD (each $\Gamma_i$ can be expanded as 
$\Gamma_i^{(0)} + \frac{\alpha_s}{\pi} \Gamma_i^{(1)} + ...$).
In addition to the leading term $\Gamma_3^{(0)}$ the following corrections we done in the 
literature for $\Gamma_{12}$:
\begin{itemize}
\item 1996: Power corrections ($\Gamma_4^{(0)}$)\cite{BBD96} turned out to be sizable.
\item 1998: NLO-QCD corrections ($\Gamma_3^{(1)}$)\cite{Beneke:1998sy} to the leading term are also
            sizeable and of conceptual importance.
\item 2000: In 1998 no lattice data for all arising matrix elements of four quark operators were
            available, the numerical update of [\refcite{Beneke:1998sy}] with lattice values was 
            given in  [\refcite{Beneke:2000cu}].
\item 2003: NLO-QCD corrections ($\Gamma_3^{(1)}$) to all CKM structures were calculated in
            [\refcite{Beneke:2003az}] and [\refcite{rome03}]. This was a relativeley small correction
            for $\Delta \Gamma$, but the dominant contribution to the semileptonic CP-asymmetries.
\item 2004: At that time all corrections to the leading term of $\Delta \Gamma$ seemed to be 
            unnatural large, this bad behaviour was summarized  in [\refcite{Lenz:2004nx}].
\item 2006: A reanalysis \cite{Lenz:2006hd} of the theoretical determination of $\Gamma_{12}$, showed
            that the above shortcommings were due to the use of an unproper operator basis with 
            large unphysical cancellations, the use of the pole $b$-quark mass and 
            the neglect of subleading CKM structures.  Taking all this into account the theoretical
            uncertainty in $\Gamma_{12}/M_{12}$ could be reduced by a facor of almost three.
\item 2007: Higher power corrections ($\Gamma_5^{(0)}$)\cite{BGP07} were estimated to be negligible.
\end{itemize}
Despite considerable efforts in the non-perturbative determination of the matrix elements
of four-quark operators entering $\Gamma_3$, see [\refcite{Lubicz:2008am}] for a recent review,
we still have a relatively limited knowledge of the decay constants, see e.g.[\refcite{Lenz:2008xt}]
for more details,
wich results in large uncertainties in $\Delta M_s$ and $\Delta \Gamma_s$ 
\footnote{$\Delta M_s$ and $\Delta \Gamma_s$ depend quadratically on $f_{B_s}$.}.
In Ref.~[\refcite{Lenz:2006hd}] we used the conservative estimate $f_{B_s} = 240 \pm 40$ MeV, while
[\refcite{Lubicz:2008am}] obtains the lattice average $f_{B_s} = 245 \pm 25$ MeV, which is very close
to the most recent QCD sum rule estimate \cite{JaminLange} $f_{B_s} = 244 \pm 21$ MeV.
In $\Gamma_{12}/M_{12}$ the decay constants cancel, and therefore $\Delta \Gamma_s / \Delta M_s$
and the semileptonic CP-asymmetries are theoretical well under control.
\\
Summarizing we can state for the theoretical uncertainties in the $B_s$ mixing quantities:
$\Delta \Gamma_s$ and $\Delta M_s$ are completely dominated by the uncertainty in the decay
constant $f_{B_s}$, while for $\Delta \Gamma_s / \Delta M_s$ and the semileptonic CP-asymmetries
conservative error estimates yield errors of about $\pm 20 \%$ \cite{Lenz:2006hd}.
\section{Conclusions}
The system of the neutral $B_s$ mesons is ideally suited for the search for  
new physics effects. In particular the standard model predicts an almost vanishing mixing phase
$\phi_s$, while we have currently some experimental 2-3$\sigma$ hints for a sizeable value of 
this phase. If this hints will be confirmed, then we have an unambiguous proof for new physics
in flavor physics.
Depending on the actual size of $\Delta$ a confirmation of the hints might already be possible
at TeVatron or at an extended $\Upsilon (5s)$ run of Belle. Precision data on $\Delta$ will be available 
from LHC and from a Super-B factory.

\section*{Acknowledgments}
I would like to thank the organizers of CAQCD 2008 for the invitation and for the financial support.



\bibliographystyle{ws-procs9x6}
\bibliography{Lenz}

\end{document}